\documentclass[fleqn,usenatbib]{mnras}

\usepackage{newtxtext,newtxmath}

\usepackage[T1]{fontenc}

\usepackage{graphicx}	
\usepackage{hyperref}  
\usepackage{subcaption} 
\usepackage{placeins}
\usepackage{arydshln} 

\hypersetup{colorlinks, urlcolor={blue}}

\title[Thermal history of the merging cluster Cygnus A]{Decoding the thermal history of the merging cluster Cygnus A}

\author[A. Majumder et al.]{
Anwesh Majumder$^{1,2}$\thanks{E-mail: a.majumder@uva.nl (AM)},
M.W. Wise$^{2,1}$,
A. Simionescu$^{2,3,4}$
and M.N. de Vries$^{5}$ \\
\\
$^{1}$Astronomical Institute ’Anton Pannekoek’, University of Amsterdam, Science Park 904, 1098 XH Amsterdam, The Netherlands\\
$^{2}$SRON, Netherlands Institute for Space Research, Niels Bohrweg 4, 2333 CA Leiden, The Netherlands\\
$^{3}$Leiden Observatory, Leiden University, PO Box 9513, 2300 RA Leiden, The Netherlands\\
$^{4}$Kavli Institute for the Physics and Mathematics of the Universe, The University of Tokyo, Kashiwa, Chiba 277-8583, Japan \\
$^{5}$Department of Physics/KIPAC, Stanford University, Stanford, CA 94305-4060, USA
}

\date{Accepted 2024 January 3. Received 2023 December 18; in original form 2023 October 13}

\pubyear{2024}

\begin{document}
\label{firstpage}
\pagerange{\pageref{firstpage}--\pageref{lastpage}}
\maketitle

\begin{abstract}
We report on a detailed spatial and spectral analysis of the large-scale X-ray emission from the merging cluster Cygnus A. We use 2.2 Ms \textit{Chandra} and 40 ks \textit{XMM-Newton} archival datasets to determine the thermodynamic properties of the intracluster gas in the merger region between the two sub-clusters in the system. These profiles exhibit temperature enhancements that imply significant heating along the merger axis. Possible sources for this heating include the shock from the ongoing merger, past activity of the powerful AGN in the core, or a combination of both. To distinguish between these scenarios, we compare the observed X-ray properties of Cygnus A with simple, spherical cluster models. These models are constructed using azimuthally averaged density and temperature profiles determined from the undisturbed regions of the cluster and folded through MARX to produce simulated \textit{Chandra} observations. The thermodynamic properties in the merger region from these simulated X-ray observations were used as a baseline for comparison with the actual observations. We identify two distinct components in the temperature structure along the merger axis, a smooth, large-scale temperature excess we attribute to the ongoing merger, and a series of peaks where the temperatures are enhanced by $0.5-2.5$ keV. If these peaks are attributable to the central AGN, the location and strength of these features imply that Cygnus A has been active for the past 300 Myr injecting a total of $\sim$10$^{62}$ erg into the merger region. This corresponds to $\sim$10\% of the energy deposited by the merger shock.

\end{abstract}




\section{Introduction}

Galaxy clusters represent the fossil record of billions of years of accumulated physics driving the formation and evolution of large-scale structure in the Universe. Shocks from mergers \citep{mar07} as well as the effects from any central active galactic nuclei (AGN; \citealt{dun06,bir12}) can heat the intracluster medium (ICM) up to temperatures of 10 keV. Such heating mechanisms are crucial to prevent catastrophic radiative cooling of the ICM and reduce the amount of gas flowing into the core \citep{bir04,raf06,dun06}. Furthermore, gas from the core can be transported out to larger radii by powerful jets from AGN forming cavities in the process (for a detailed review, see \citealt{mcn07}). Such cavities subsequently rise buoyantly through the ICM transporting energy and metals \citep{mcn07,san05,roe07}. Thus, better constraints on the energy input into the ICM from both mergers and AGN feedback is essential to understand the evolution of clusters over cosmic time.

Although the key physics involved in these two mechanisms is broadly known, the details are poorly understood. For example, the fraction of energy released by mergers that is deposited into the ICM is not well constrained. Furthermore, any additional heating of the ICM by a central AGN complicates our understanding of the relative contribution of these two processes to the overall energy budget. While some past studies have tried to tackle the problem of energy transport due to AGN from the core to the ICM \citep{rus02,rus04,san07,fab17}, others have focused on energy flows from outskirts to the core due to mergers (e.g., \citealt{chu03,bru12}). There is, however, a lack of suitable targets where these two important processes can be studied simultaneously to understand their relative importance. 

The nearby powerful FRII-class radio galaxy Cygnus A \citep{fan74} is an excellent object to study these two physical processes in detail. At a redshift of $z$ = 0.0561, it is $10^3$ times brighter than any other AGN at similar distances \citep{sto96,car96} and embedded in a large reservoir of X-ray emitting cluster gas extending over a Mpc from the center \citep{arn84,rey96}. The subcluster containing the radio source Cygnus A is furthermore undergoing a major merger at the largest scales that is driving a shock in the outer cluster atmosphere \citep{owe97,mar99,smi02,led05}. Due to the proximity and extreme brightness of Cygnus A, it is possible resolve X-ray features from kpc scales near the core out to Mpc scales. Hence, it is possible to investigate signatures of AGN feedback, structures associated with the merger, as well as any possible interaction between the two. This makes Cygnus A an ideal laboratory to investigate both AGN feedback and merger effects on the ICM in unprecedented detail.

We use 2.2 Ms of archival \textit{Chandra} ACIS data to create thermodynamic profiles of this merging system. The high spatial resolution of \textit{Chandra} allows us to create detailed temperature and density profiles along the merger direction that reveal a very hot and complex ICM. In order to separate signatures of heating from the underlying temperature structure of the undisturbed ICM, we construct a model for the system assuming hydrostatic equilibrium. Any deviations from this quiescent model can be attributed to other physical mechanisms such as feedback from the AGN or the merger. Furthermore, we also make use of a 40 ks archival \textit{XMM-Newton} exposure of Cygnus A to independently verify the detected heating signatures. By modeling the various contributions to the detected level of heating, we can provide new constraints on the effectiveness of both processes in heating the ICM.

We begin in \S\ref{Data} with a description of the basic data reduction used to create the event files and mosaic images. Our paper carefully takes into account any Galactic foreground emission and this is subsequently discussed. This discussion is followed by a general description of the large-scale X-ray morphology in \S\ref{lss}. Thermodynamic profiles of the system are presented in \S\ref{profiles} for both the merger and non-merger regions of the system. The construction of our model for the undisturbed cluster system, the process to create the simulated datasets, and the subsequent derived model thermodynamic profiles are described in \S\ref{Simulation}. An analysis of the excess observed temperature and energy is shown in \S\ref{excess_energy}. The implications of our analysis are summarised in \S\ref{summary}.

In this paper we have assumed a $\Lambda$CDM cosmology with $H_0$= 70 km s$^{-1}$ Mpc$^{-1}$, $\Omega_m$ = 0.3, $\Omega_{\Lambda}$ = 0.7, yielding a linear scale of 65.3 kpc per arcminute at Cygnus A's redshift. All errors reported in this paper are of $1\sigma$ significance.

\section{Observations and data reduction} \label{Data}

\subsection{\textit{Chandra}} \label{Chandra_repro}

We have reprocessed and analysed all existing observations of Cygnus A in the Chandra archive. These include deep observations from a multiwavelength observational campaign by \cite{wise14}. We have, however, excluded ObsID 359 and 1707 from our analysis as they were taken to study photon pileup from the nucleus with short CCD readout times. The rest of the 71 ObsIDs have a combined exposure time of 2.199 Ms. 

We use standard reprocessing of each individual observation using \texttt{CIAO} 4.13 \citep{fru06} and \texttt{CALDB} 4.9.4. All ObsIDs were examined for contamination due to strong background flares by extracting light curves from a chip not containing the bulk of the cluster emission. Only two ObsIDs (5830 and 6226) had a mild signature of flares and the affected time intervals ($\sim 1.5$ ks) were removed. The combined exposure after all the filtering is 2.198 Ms. For VFAINT mode observations, we filter background particles in the event files during reprocessing by setting \texttt{check\_vf\_pha=yes}.

In our analysis, we treat the non-X-ray background (NXB) and X-ray background (XRB) for Cygnus A separately. We discuss modeling the X-ray background in \S\ref{XRB}. To determine the instrumental background, we use the ``stowed" ACIS event files which are available in \texttt{CALDB}. This background was measured when the instrument was not exposed to the sky. We created background event files for each individual ObsID by applying the correct gains to the ``stowed" files followed by reprojecting them to the same tangent plane. For VFAINT mode observations, we filter the background files further by only selecting \texttt{status=0} events. 

\begin{table}
\caption{X-ray foreground components constrained by the RASS spectrum. The temperature of `low Galactic latitude stars' component was held fixed during the fit. The normalizations are scaled to a 1 arcmin$^2$ area.}
\centering
\begin{tabular}{c c c} 
 \hline
 \hline
  & Flux (0.1-2.4 keV) & $kT$ \\ 
  & erg s$^{-1}$ cm$^{-2}$ & keV \\
 \hline
Galactic halo& (6 $\pm$ 2) $\times$ $10^{-15}$& 0.18 $\pm$ 0.03\\ 
Low Galactic latitude stars& (1.3 $\pm$ 0.3) $\times$ $10^{-15}$& 0.6\\ 
Local hot bubble& (1.4 $\pm$ 0.2) $\times$ $10^{-16}$& 0.22 $\pm$ 0.15\\
 \hline
\end{tabular}
\label{fit_table}
\end{table}

\subsection{\textit{XMM-Newton}}

Cygnus A has been observed by \textit{XMM-Newton} for a total of 40 ks (ObsIDs: 0302800101 and 0302800201) in full-frame mode with medium filter. Both ObsIDs were reduced using the \textit{XMM-Newton} Science Analysis System (SAS) v18.0.0. We use standard processing to obtain MOS and PN event files from the observation data files. Both the ObsIDs were observed in Full-Frame mode and the PN out-of-time events were taken into account as per standard analysis\footnote{\href{https://www.cosmos.esa.int/web/xmm-newton/sas-thread-epic-oot}{{https://www.cosmos.esa.int/web/xmm-newton/sas-thread-epic-oot}}}.

We remove contamination due to soft-proton flares by building good time intervals. We fit gaussians to the count rate histogram and reject all time intervals with count rate more than $\mu + 2\sigma$, where $\mu$ is the mean and $\sigma$ is the standard deviation. We find the rejected time interval to be negligible.

For the \textit{XMM-Newton} non-X-ray background, we choose the filter wheel closed event files of XMM-Newton\footnote{\href{https://www.cosmos.esa.int/web/xmm-newton/filter-closed}{https://www.cosmos.esa.int/web/xmm-newton/filter-closed}}. These event files were then reprojected to match sky coordinates of individual ObsIDs. The event files were further filtered to remove flares in an identical manner as the observed data.  

\subsection{Modeling the X-ray background} \label{XRB}

The Cygnus A event files include a significant contribution at soft energies from Galactic emission because of its proximity to the Galactic plane (l=76.19 b=+05.76). This Galactic soft emission was modeled using ROSAT all-sky survey (RASS) spectra generated by the X-ray background tool\footnote{\href{https://heasarc.gsfc.nasa.gov/cgi-bin/Tools/xraybg/xraybg.pl}{https://heasarc.gsfc.nasa.gov/cgi-bin/Tools/xraybg/xraybg.pl}} \citep{sab19}. The RASS spectra were selected from a region with minimal cluster emission to maximize the detected counts from the actual X-ray background. We select an annular region centered around Cygnus A with an inner radius of $1.5$R$_{200}$ and outer radius of $1.5$R$_{200}$ + 1$^{\circ}$. The spectra was extracted from this region and analyzed in XSPEC to estimate the contribution from Galactic soft X-ray emission. This procedure yields 16869 counts in the energy band 0.1$-$2.4 keV.

We fit a \texttt{tbabs(apec+apec+powerlaw)+apec} model to this spectrum to model the Galactic halo, low Galactic latitude stars, Galactic non-thermal emission and the local hot bubble, respectively. The first three components are absorbed by the Galactic neutral hydrogen component while the local hot bubble is unabsorbed. We set the abundance value for each \texttt{apec} component to proto-solar values while the redshift was set to zero. The powerlaw photon index of Galactic non-thermal emission was set to 1.46 and the powerlaw norm was set to $8.88 \times 10^{-7}$ photons\textbf{/s}/keV/cm$^2$ at 1 keV \citep{sno08}. As can be seen in Fig. \ref{fig:chandra_cyga_properties}, the XRB is much fainter than the ICM signal at all radii considered here, therefore any uncertainties in the background model do not impact our results significantly. The hydrogen column density of Cygnus A was calculated using the method of \cite{wil13}\footnote{\href{https://www.swift.ac.uk/analysis/nhtot/}{https://www.swift.ac.uk/analysis/nhtot/}} which takes both atomic and molecular hydrogen into account. The weighted effective column density is $n_H = 4.14 \times 10^{21}$ atoms cm$^{-2}$. We used the latest available abundance model \texttt{asplund} \citep{aspl09} in \texttt{XSPEC v12.10.1f}.  We chose this model as the \texttt{He/H} ratio\footnote{\href{https://heasarc.gsfc.nasa.gov/xanadu/xspec/manual/node116.html}{https://heasarc.gsfc.nasa.gov/xanadu/xspec/manual/node116.html}} is close to the primordial value from the Big Bang \citep{ste07}. The fit parameters of the Galactic soft X-ray spectrum are reported in Table \ref{fit_table}. The quality of the fit is \texttt{$\chi^2/$dof = 0.46/2}.

In all subsequent spectral analysis, we scale the normalizations of the soft energy model components to the area of the corresponding extraction regions. The derived parameters for the soft X-ray background model are then held fixed while fitting the cluster emission.  

\begin{figure}
    \centering
    \includegraphics[width=\columnwidth]{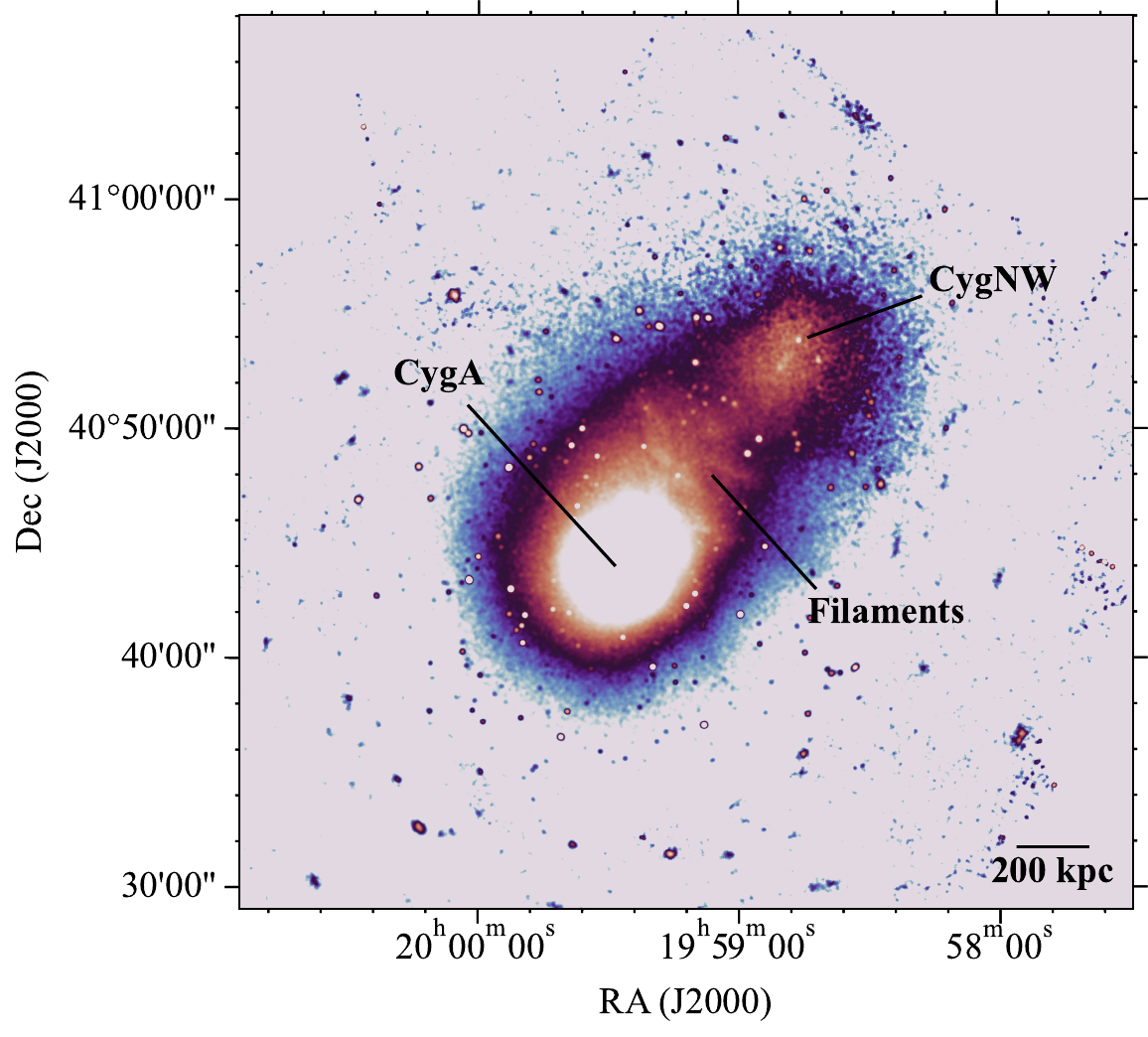}
    \caption{\small Exposure-corrected, background-subtracted 0.7-7.0 keV surface brightness mosaic for the full field of view covered by the existing 2.2 Ms Chandra data. The field measures $0.8 \times 0.9$ deg$^2$ and has been smoothed with a $\sigma = 10''$ Gaussian to enhance the appearance of the diffuse emission. The merging subclusters, CygA and CygNW, are visible along with a filamentary structure of material in the interstitial region between them. The bright X-ray emission associated with the Cygnus A radio galaxy is clearly visible in the centre of the CygA subcluster. These features are discussed in \S\ref{lss} and \S\ref{profiles}.}
    \label{fig:chandra_image}
\end{figure}

\subsection{Image processing}

To create mosaiced images, all \textit{Chandra} event files were reprojected to a common tangent plane. Count images and exposure maps for each reprojected event file were created in the $0.7-7.0$ keV band and combined to produce total image and exposure maps. The combined image was inspected for point sources using the script \texttt{wavdetect}. The detected point sources were visually confirmed before being removed from the combined image and from each event file. 

Background images from individual background files were created and scaled to match the $9.0-12.0$ keV emission in the observed event files. This energy band is suitable as it is dominated by NXB due to the negligible ACIS effective area. The resulting background-subtracted, exposure-corrected mosaic for the full field of view around Cygnus A in the energy range $0.7-7.0$ keV is shown in Figure \ref{fig:chandra_image}. These combined counts images, background images, and exposure maps were also used to determine the surface brightness profiles in \S\ref{profiles}. 

We use the XMM-Newton data only for spectral analysis. Hence, no images were constructed for both MOS and PN datasets.

\section{Large scale properties} \label{lss}

Figure  \ref{fig:chandra_image} shows a smoothed, exposure-corrected, background subtracted X-ray mosaic of the \textit{Chandra} data in the $0.7-7.0$ keV band. It is clear from the image that the large scale emission exhibits complex morphology and the diffuse emission extends to well over a Mpc from the center of Cygnus A. This extended emission has been observed previously \citep{arn84, rey96,smi02}, but here we can see it clearly consists of two distinct subclusters in the Cygnus A merging system. In the proceeding analysis, we shall designate the subcluster containing the Cygnus A radio source as CygA and the other subcluster to the NW as Cyg NW. 

The X-ray surface brightness of the CygA subcluster is highly peaked around the radio galaxy itself, so we assume the centroid of the subcluster to be co-located with the central AGN (RA=19:59:28.356, Dec=+40:44:02.097; \citealt{fey04}). On the other hand, the surface brightness of the CygNW subcluster is significantly fainter with an extended central concentration. To estimate the centroid of the emission from CygNW, we take the median position of all events within a 1 arcmin radius centered on the apparent peak of the surface brightness. This procedure gives us a centroid of (RA=19:58:47.313, Dec=40:53:17.280) for CygNW. Based on the choice of centroids, the two subclusters are separated by $\sim$12 arcmin or 784 kpc, oriented along an axis with a position angle of $140^{\circ}$ as measured East from North. The merger axis has an offset of $\sim$30$^{\circ}$ from the major axis of the Cygnus A radio galaxy as defined by the line connecting the radio hotspots.

The large scale morphology suggests that the two subclusters are undergoing a merger. This assertion is supported by previous studies \citep{owe97,mar98,mar99,led05} which argue that the system is in the early stages of a merger $\sim$0.5 Gyr prior to the initial core passage. The region in between the two subclusters shows enhanced surface brightness and contains filamentary structure in the merger region at radii between $200-400$ kpc from the center of CygA. The position of these filaments is consistent with the location of a region of enhanced termperature seen in lower angular resolution ASCA data \citep{mar99,sar13}. In the next sections, we will investigate these features in more detail. 

\subsection{Region selection} \label{regions}

\begin{figure}
    \centering
    \includegraphics[height=8.25in,keepaspectratio]{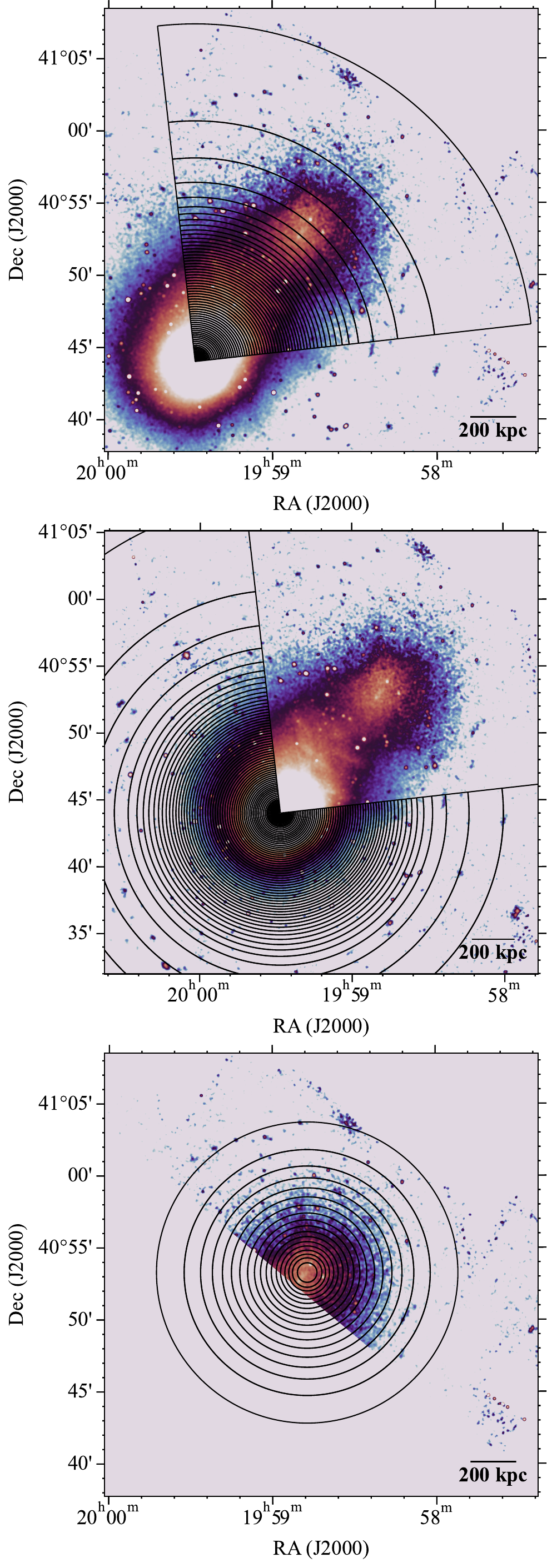}
    \caption{\small The regions used for analysing the merger side of CygA subcluster are shown. \emph{Middle:} The regions used for analysing the non-merger side of CygA subcluster are shown. \emph{Bottom:} The regions used for analysing the non-merger side of CygNW subcluster are shown. All three images have been exposure-corrected and background-subtracted in the 0.7-7.0 keV band. They have been smoothed with a $\sigma = 10''$ Gaussian to enhance the appearance of the diffuse emission.}
    \label{fig:regions}
\end{figure}

In order to clearly isolate the physical processes occurring in the merger region, we have separated the cluster into two annular sectors, one centered on the merger axis and the other containing the remainder of the undisturbed cluster emission. We define the merger region by choosing a $90^{\circ}$ wedge centered on the merger axis between the two subclusters (see Figure \ref{fig:regions}). This region encloses the interaction region and extends from the center of the CygA subcluster to the outskirts of the CygNW subcluster. This region was adaptively divided into annular bins centered on the CygA subcluster such that each bin has at least 50,000 counts (SNR $\sim 223$). This count rate was chosen to ensure the error on the fitted temperature is less than 5\% in the hot interstitial region. This choice gives us 70 distinct annular bins with a bin width that ranges from 2 arcsec near the core to 5 arcsec across the merger region.   

The non-merger region for the CygA subcluster was defined by excluding the merger region from the event files. This region is shown in Figure \ref{fig:regions}. Spectra for the non-merger portion of the cluster were extracted using the same radial binning as the merger region to allow direct comparison between the thermodynamic profiles for both regions. The non-merger region for the CygNW subcluster was chosen in a way to minimise the contribution of emission from CygA. This region is also shown in Figure \ref{fig:regions}. We then adaptively divided this region into annular bins centered on the CygNW subcluster such that each bin has at least 10,000 counts (SNR $\sim 100$).

Due to the low exposure of the \textit{XMM-Newton} data, we divided the merger region into annular bins such that each bin has at least $\sim$1600 counts (SNR $\sim40$). This gives us an error on the fitted temperature of $\sim$10\% and results in a bin width of $\sim
$10 arcsec at the core and $\sim$20-30 arcsec in the merger region. We only analyzed the \textit{XMM-Newton} data in the merger region as the exposure is too low to be used in the fainter non-merger region.

\begin{figure*}
    \centering
    \includegraphics[width=8in,height=8in,keepaspectratio]{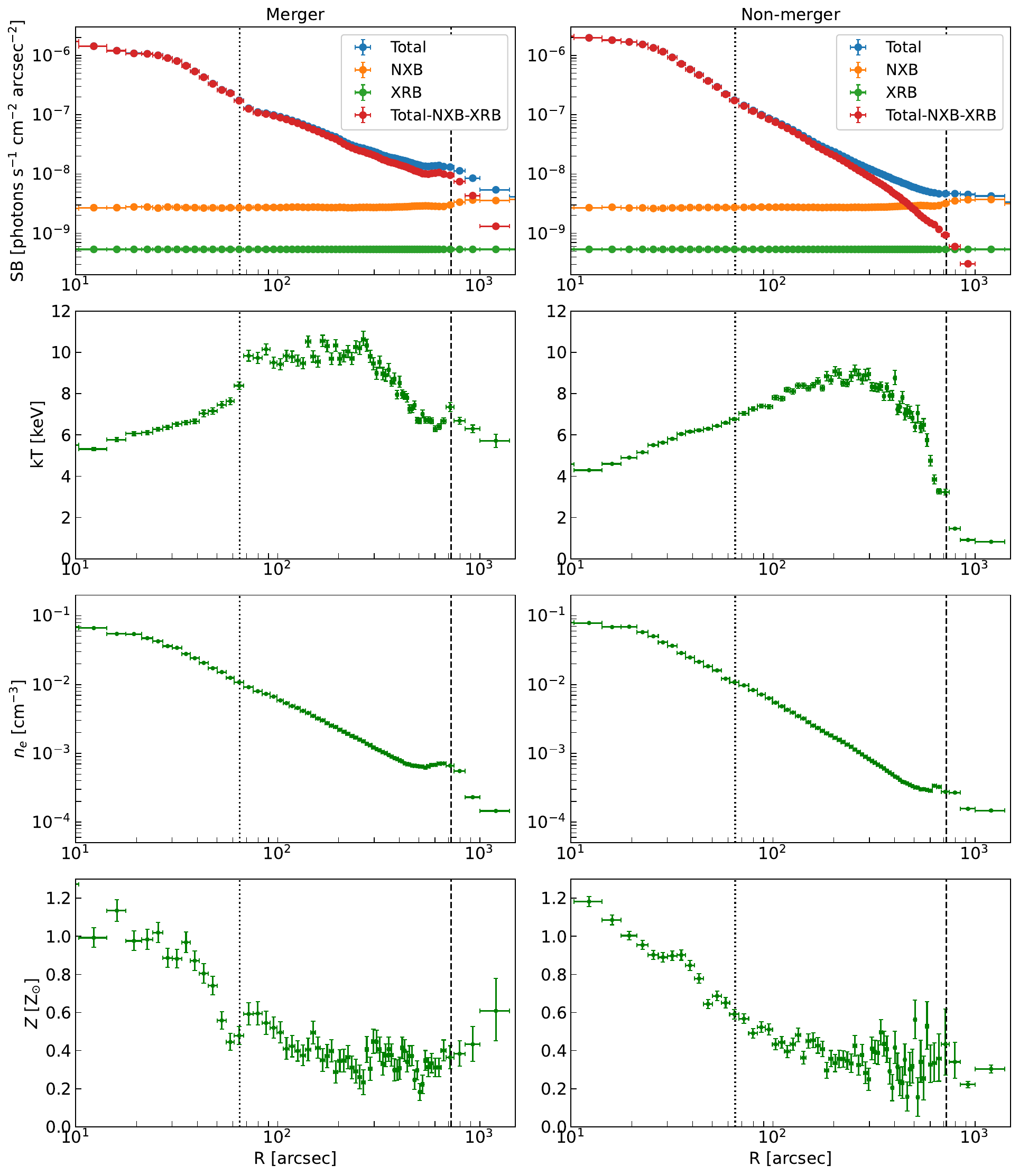}
    \caption{\small The left panels correspond to profiles along the merger region while the right panels correspond to profiles along the non-merger region of CygA. \emph{First row:} Surface brightness profile as a function of projected radius. \emph{Second row}: Projected gas temperature as a function of projected radius. \emph{Third row}: Pseudo-deprojected electron density as a function of projected radius. \emph{Fourth row}: Metallicity profile as a function of projected radius. The dotted and dashed lines show the location of cocoon shock in CygA and the peak of the CygNW subcluster, respectively.}
    \label{fig:chandra_cyga_properties}
\end{figure*}

\begin{figure*}
    \centering
    \includegraphics[width=5in,height=5in, keepaspectratio]{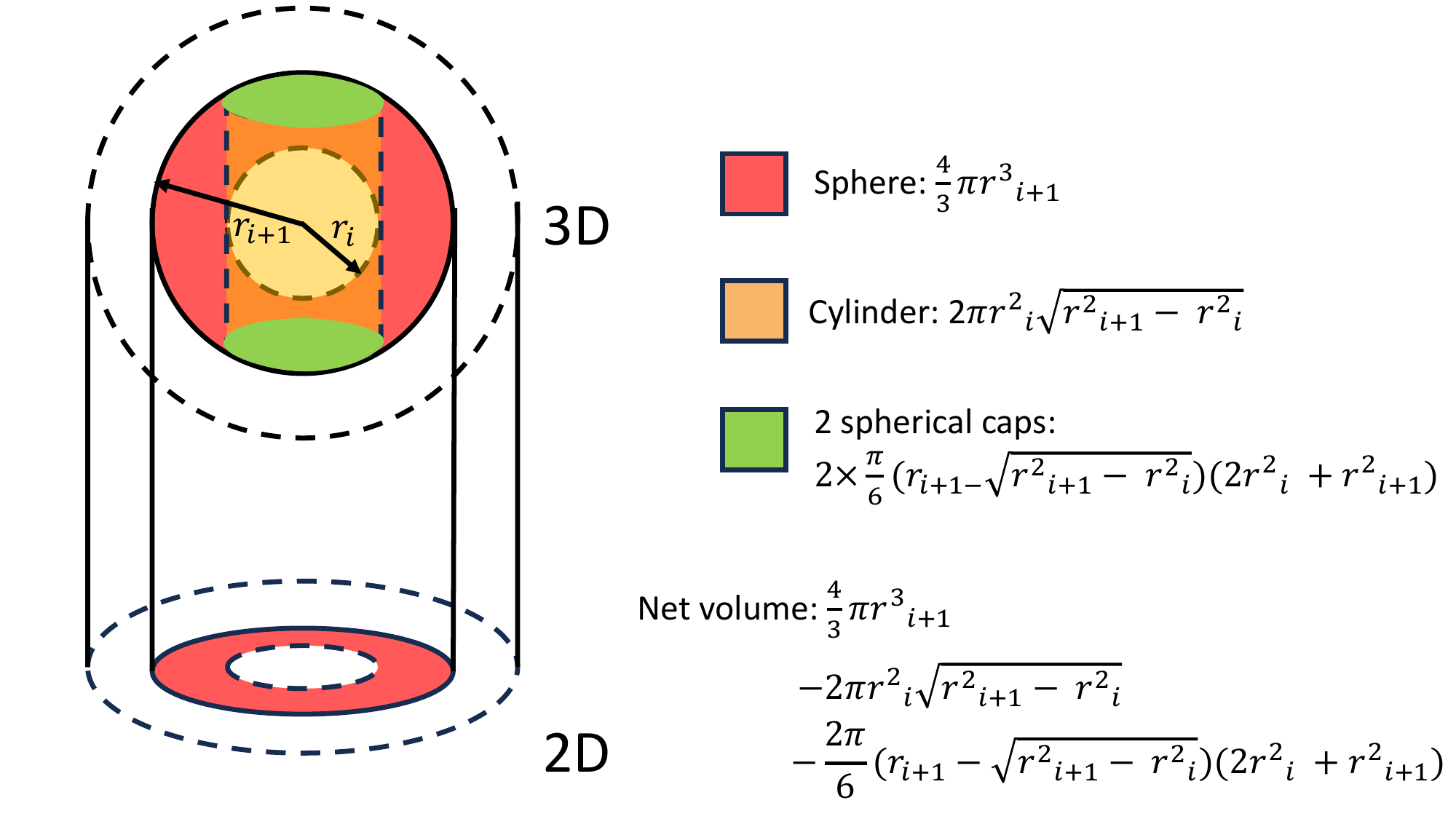}
    \caption{\small Schematic diagram for the calculation of deprojected volumes. The emission from the 3D spherical shell (shown in red) excluding the cylinder and spherical caps is detected on the 2D annulus (also shown in red). We assume that the emission from the outer spherical shell (shown with dashed line) does not contribute to the same annulus. This approximation is based on the fact that the ICM is less dense in the outer shell.}
    \label{fig:volume}
\end{figure*}

\subsection{Spectral analysis}

For each \textit{Chandra} OBSID, individual source, background spectra, and response files within a given extraction region were obtained using the script \texttt{specextract}. The \texttt{BACKSCAL} header of each of the instrumental background spectrum was then scaled so that the source and background count rate match in the $9-12$ keV band. For \textit{XMM-Newton}, the XMM-SAS task \texttt{evselect} was used to extract both source and background spectra with \texttt{FLAG==0}. We only select single to quadruple events in MOS (\texttt{pattern <= 12}) and single events in PN (\texttt{pattern == 0}). The redistribution matrix file (RMF) and ancillary response file (ARF) for each spectra were generated with the tasks \texttt{rmfgen} and \texttt{arfgen}. A user defined detector map was provided for each of the tasks to better model spatial variation of the source. The background spectra were then scaled, as discussed in \S\ref{Chandra_repro}, and used in all subsequent analysis.  

For both \textit{Chandra} and \textit{XMM-Newton}, we fit spectra from all ObsIDs for a given extraction region simultaneously using a \texttt{tbabs(apec)} model in \texttt{XSPEC} along with the Galactic soft emission model described in \S\ref{XRB}. We use \texttt{asplund} abundance model in our fitting and use \texttt{cstat} in all our analysis. This setup gives us good spectral fits with the value of C-statistic comparable to the number of degrees of freedom. For \textit{Chandra} data, the spectral fits had \texttt{cstat/dof = 15688/13789} in the outermost bin of the merger region and \texttt{cstat/dof = 24727/21547} in the outermost bin of the non-merger region. For \textit{XMM-Newton} data on the other hand, the spectral fits had \texttt{cstat/dof = 245/253} in the outermost bin of the merger region. 

\section{Thermodynamic profiles} \label{profiles}
\subsection{Surface brightness}

The radial surface brightness profiles in the merger and non-merger directions of Cyg A were extracted using the same radial grid used for the spectral extractions. A comparison of these profiles is shown in Figure \ref{fig:chandra_cyga_properties}. Total contributions from the NXB and XRB are shown along with the resulting background subtracted profile. The Galactic foreground is calculated in the $0.7-7.0$ keV band by using our Galactic emission model normalized assuming \textit{Chandra} on-axis effective area and response. The surface brightness of this component is assumed to be constant across the whole field of view. At $\sim$50 arcsec, we see a sharp jump in temperature as well as a sharp discontinuity in the surface brightness profile. These features are the result of the well-studied cocoon shock \citep{beg89,car_artcile_96,cla97,wil06,sni18} being driven into the surrounding ICM by the central AGN (see Figure \ref{fig:chandra_image}). We also see  see a bump in all the profiles at $\sim800$ arcsec because of the presence of CygNW. There are also some fluctuations in the profile along the merger region which correspond to the location of the surface brightness features seen in Figure \ref{fig:chandra_image}. In contrast, the surface brightness profile along the non-merger direction (Figure \ref{fig:chandra_cyga_properties}) is relatively smooth. This smoothness suggests that this region is consistent with being in hydrostatic equilibrium and representative of the undisturbed cluster atmosphere on larger scales.

\begin{figure*}
    \centering
    \includegraphics[width=7in,height=7in, keepaspectratio]{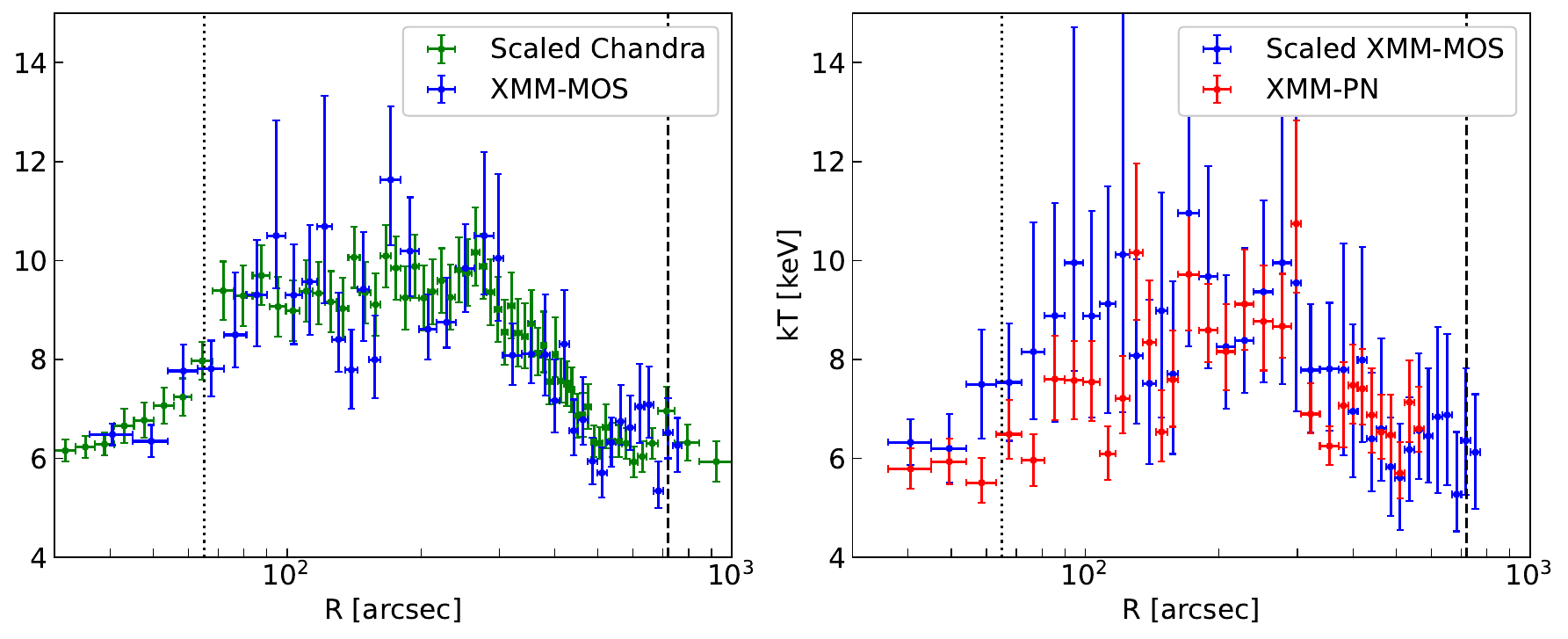}
    \caption{\small \emph{Left}: Projected temperature profile across the CygA-CygNW merger region based on spectral fits to 40 ks observations from XMM-MOS. The scaled \textit{Chandra} profile has been shown for comparison. \emph{Right}: Projected temperature profile across the CygA-CygNW merger region based on spectral fits to 40 ks observations from XMM-PN. The MOS temperature has been scaled as well to compare with the XMM-PN profile. The dotted line shows the approximate location of the cocoon shock while the dashed line shows the location of CygNW subcluster.}
    \label{fig:chandra_xmm_comp}
\end{figure*}

\subsection{Temperature} \label{temperature}

The temperature profiles along the merger and non-merger directions of CygA are shown in Figure \ref{fig:chandra_cyga_properties}. It is immediately obvious that the gas on the merger side is hotter everywhere by $0.5-2.5$ keV compared to the non-merger side. As mentioned above, we also notice the presence of a clear jump in temperature at $60-70$ arcsec in the merger profile that is consistent with the cocoon shock (see \citealt{sni18}). 

It is interesting to note that there are a number of features present in the merger profile between $\sim100-400$ arcsec. The temperature is fluctuating in this region by $0.5-1.0$ keV on top of the underlying profile. We estimate and analyze these fluctuations in detail in \S\ref{excess_energy}. The non-merger side however does not show any of these features and has a smooth profile. These fluctuations form a series of ripples. The origin of such temperature features is unclear but they correspond to the filamentary structures discussed in \S\ref{lss} and seen in Figure \ref{fig:chandra_image}. The implications of such temperature variation is explored in \S\ref{excess_energy}.      

\subsection{Density} \label{density}

Using the normalization of the \texttt{APEC} model, we can estimate the gas density by using the following equation:

\begin{equation}
    n_en_p = 4\pi D_A^2 (1+z)^2 10^{14} V^{-1} N_{\textrm{\texttt{APEC}}}
\end{equation}
where $D_A$ is the angular diameter distance to the source in cm, $z$ is the redshift, V is the volume of the region along the line of sight, and $N_{\textrm{\texttt{APEC}}}$ is the normalization of the \texttt{APEC} model obtained from the spectral fit. The electron to proton density ratio is 1.18 for the assumed \texttt{ASPLUND} abundance profile ($n_e \approx 1.18n_p$).  We estimate the volume, $V$, as the intersection between a spherical shell and a cyllinder, with radii corresponding to the projected 2D annulus in the plane of the sky. Each of the projected annular bins along the $90^{\circ}$ wedge has a line of sight volume of (see Figure \ref{fig:volume}):

\begin{align} \label{volume}
    V = \frac{1}{4} \Big(\frac{4\pi}{3}r_{i+1}^3 - 2\pi r_i^2\sqrt{r_{i+1}^2-r_{i}^2} -  \frac{2\pi}{6}\Big(r_{i+1} - \sqrt{r_{i+1}^2-r_{i}^2}\Big)\times \nonumber\\
    \Big(2r_i^2 + r_{i+1}^2\Big)\Big) 
\end{align}
where $r_i$ is inner radius of the annular bin and $r_{i+1}$ is the outer radius as measured from the centre of Cygnus A. This volume is an approximation assuming all the emission in the annulus comes from the spherical shell between $r_i$ and $r_{i+1}$. We neglect emission from larger radii that lie along the line of sight behind and in front of this spherical shell.

The pseudo-deprojected electron density profiles along both merger and non-merger directions of Cyg A are shown in Figure \ref{fig:chandra_cyga_properties}. The presence of the CygNW subcluster is clearly visible at $\sim$800 arcsec in the merger profile. The non-merger profile, shown in Figure \ref{fig:chandra_cyga_properties}, has a slight bump at the position of CygNW. While we have taken care to exclude CygNW, some minor contribution could remain due to the very extended nature of CygNW. We also note that the actual geometry near CygNW likely differs from the spherical geometry assumed in this section. The exact volume along the line of sight in this region is likely an intersection between a sphere centered on CygA and another sphere centered on CygNW. The volume expression used in Eq. \ref{volume}, however, does not take into account such a complicated geometry. As such, systematic uncertainties can be associated with density calculations near CygNW.

\subsection{Abundance}

We show the iron abundance profile in the merger and non-merger direction of Cyg A again in Figure \ref{fig:chandra_cyga_properties}. There is a sharp drop in abundance at $\sim$60 arcsec, consistent with the location of the cocoon shock. From previous work, it is known that there is significant non-thermal emission around the cocoon shock \citep{sni18} and it is likely that fitting spectra from regions around the shock can lead to lower abundance measurements, due to imprecision in estimating the thermal continuum. Alternatively, the temperature of the ICM across the shock boundary is different and fitting such multi-temperature gas spectra with a single temperature \texttt{APEC} model can lead to a drop in abundance around the shock. Near the core, the abundance profile is peaked which is expected for a cool-core cluster like Cygnus A \citep{deg01,mer17}. Although the errors on the profile become large at the outskirts, it is clear that beyond 200 arcsec, the abundance on both the merger and non-merger sides is similar and flat. This supports the early-enrichment scenario where metal enrichment occurs during the proto-cluster phase \citep{wer13}. Overall, the abundance profile of Cygnus A is similar in shape to previous statistical studies of abundance profiles \citep{mer17}.

\subsection{Comparison of temperature profile with \textit{XMM-Newton} data} \label{comparison}

The temperature profile of Cygnus A reveals features consistent with the location of filamentary structure on the merger side as well as an overall enhancement due to the merger. As a further check, we compare our \textit{Chandra} merger temperature profile with XMM-MOS and PN data as it provides an independent verification that the observed features are physical and not due to issues with the \textit{Chandra} data extraction or analysis. For this comparison, we must take into account the systematic differences in fitted temperatures determined with MOS, PN and ACIS spectra. Such differences occur because of cross-calibration uncertainties of the effective area between different detectors. Hence, we need to scale the \textit{Chandra} and PN temperature profile before comparing it with the XMM-MOS profile. The scaling relationship is obtained from \cite{sch15}:

\begin{equation}
    \log_{10}\frac{kT_{I_{Y,\textrm{band}}}}{\textrm{1 keV}} = a \times \log_{10}\frac{kT_{I_{X,\textrm{band}}}}{\textrm{1 keV}} + b
\end{equation}
where the values of $a$ and $b$ depend on the type of detector and energy band. For the hard band and ACIS-EMOS detectors, the best-fit values are: $a = 1.028$, $b = -0.048$. The values for EMOS-EPN detectors in the same band, on the other hand, are: $a = 0.940$, $b = 0.038$. We used the hard band values for the Cygnus A merger region as it is very hot ($\geq 6$ keV).

Using these values, the scaled \textit{Chandra} profile is calculated and shown in Fig. \ref{fig:chandra_xmm_comp}. The XMM-MOS profile is also shown in the figure for comparison. The overall large scale temperature profile from both \textit{Chandra} and MOS are in agreement with each other. All the temperature features discussed in \S\ref{temperature} are also visible in the MOS profile. Although there are significant uncertainties, it is evident that the ripples are situated in a comparable location to those in \textit{Chandra}. We further show the XMM-MOS profile overlaid on XMM-PN profile in Fig. \ref{fig:chandra_xmm_comp}. The PN temperature profile is consistent with MOS as well. This gives us an independent confirmation that the features are real and not due to data analysis techniques or artifacts of one particular instrument. 

\begin{figure*}
    \centering
    \includegraphics[width=8in,height=8in, keepaspectratio]{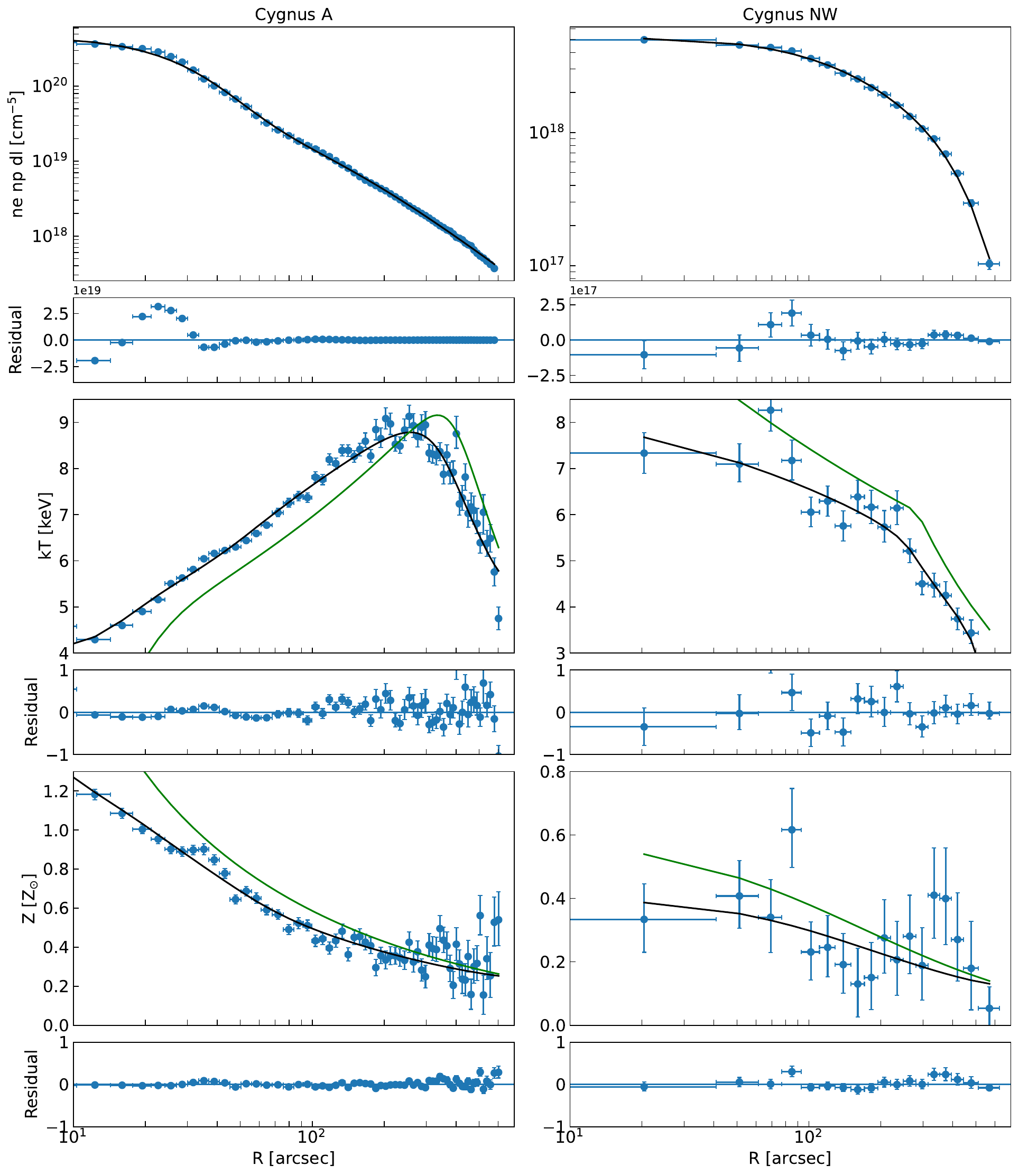}
    \caption{\small The left panels show profiles along the non-merger direction of CygA while the right panels show profiles along the non-merger direction of CygNW. \emph{First row:} Projected density profiles as a function of projected radius. \emph{Second row:} Projected gas temperature as a function of projected radius. \emph{Third row:} Projected metallicity profile as a function of projected radius. The fitted, projected profiles discussed in \S\ref{thermodynamic_profiles} are shown in black. The corresponding 3D profiles are shown in green. The residuals for each fit are shown immediately below each figure.}
    \label{fig:marx_fit}
\end{figure*}

\section{Simulation setup} \label{Simulation}

The source of heating of the ICM along the merger direction is not obvious. As the system is undergoing a merger and there is AGN activity at the center of Cygnus A, it is possible that the ICM has been heated by both of these two processes. The central objective of our work is to study how much energy is added to the ICM by such processes and what is their relative contribution. Thus, it is imperative to compare the CygA-CygNW merging system with a model where the ICM is neither being heated due to interactions of the two subclusters nor by the central AGN. We achieve this by building a model of both CygA and CygNW subclusters based on their locations and physical properties as determined outside the merger region. We use this model to create simulated \textit{Chandra} datasets which we then analyze in the same way as the actual data. The temperature difference between the actual data and the model will be the excess heating due to the combined effects of the merger and the AGN. The creation of the simulated \textit{Chandra} data and subsequent analysis is described in more detail in the following sections.

\subsection{Source model} \label{PyxSIM}

We construct a simple spherical model of a cluster using \texttt{pyXSIM}\footnote{\href{https://hea-www.cfa.harvard.edu/~jzuhone/pyxsim/index.html}{PyXSIM documentation}}, an implementation of the \texttt{PHOX} algorithm, designed to create mock X-ray observations \citep{bif12, bif13}. We set up a spatial grid of varying resolution, with finer grids at the center of the subclusters and coarser grids at outskirts. The resolution is chosen such that the pixels are no larger than 1/3rd of radial size of the spectral regions (as discussed in \S\ref{regions}). This is achieved using an adaptive mesh refinement (AMR) grid \citep{ber89} that is available inside the \texttt{yt} module\footnote{\href{https://yt-project.org}{yt project webpage}} \citep{tur11}. The density, temperature and metallicity in each cell is set according to the following profiles \citep{vik06,mer17}:

\begin{equation}
n_en_p(r) = n_0^2 \frac{(r/r_c)^{-\alpha}}{(1+r^2/r_c^2)^{3\beta-\alpha/2}} \frac{1}{(1+r^{\gamma}/r_s^{\gamma})^{\epsilon/\gamma}} + \frac{n_{02}^2}{(1+r^2/r_{c2}^2)^{3\beta_2}} \label{equation_nenp}
\end{equation}

\begin{equation}
T(r) =  \frac{T_0(x+T_{min}/T_0)}{x+1} \frac{(r/r_t)^{-a}}{(1+(r/r_t)^b)^{c/b}}  \label{equation_kT} 
\end{equation}

\begin{equation}
Z(r) = A(r/r_{500} - B)^C \label{equation_z}
\end{equation}
where $x = (r/r_{cool})^{a_{cool}}$ and r$_{500}$ is the distance at which mean density of the cluster is 500 times the critical density of the Universe. 
We set density, temperature and abundance to zero beyond $R_{200}$ which we set as our simulation boundary for each cluster. Using values from \cite{hal19}, we set the boundary of CygA and CygNW to 1831 and 1609 kpc respectively. 

Next, we create photons from each cell in this sphere according to an \texttt{APEC} model \citep{smi01} based on AtomDB \citep{fos18} in the energy range $0.7-7.0$ keV. These photons are then projected onto the sky plane and absorbed with a hydrogen column density of $4.14 \times 10^{21}$ atoms cm$^{-2}$. We use the XSPEC abundance model \texttt{asplund} and hydrogen column model \texttt{wabs} for our simulation. Finally, the projected photons are written to a SIMPUT\footnote{\href{http://hea-www.harvard.edu/heasarc/formats/simput-1.1.0.pdf}{SIMPUT documentation}} file as a photon list. This SIMPUT file is then used as input to the \texttt{MARX} instrument simulator for creating realistic \textit{Chandra} observations.

\begin{figure}
    \centering
    \includegraphics[width=\columnwidth]{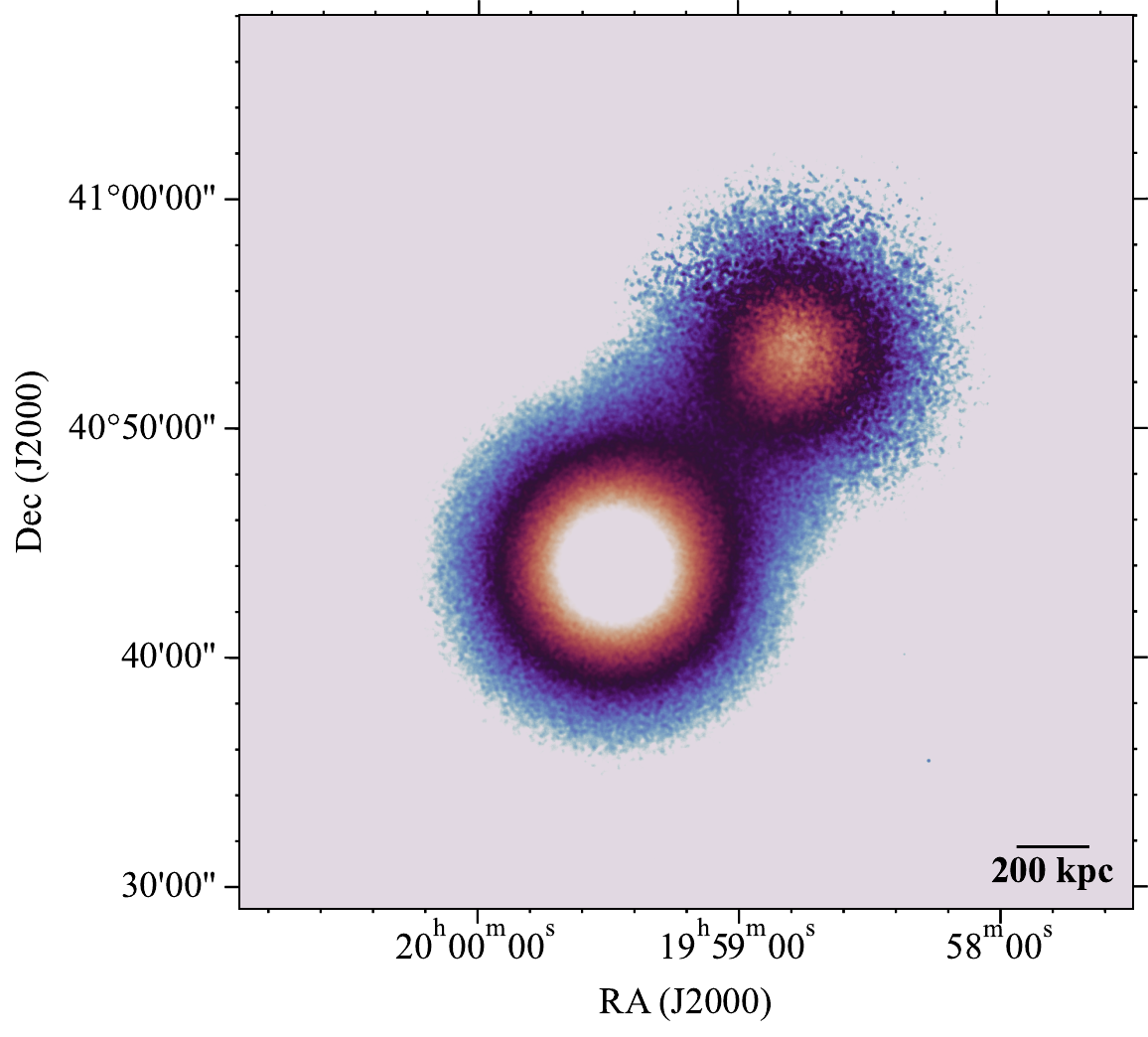}
    \caption{\small Exposure-corrected 0.7-7.0 keV surface brightness mosaic for the 2.2 Ms MARX simulation. The field of view is the same as that covered by the observed \textit{Chandra} mosaic (measuring $0.8 \times 0.9$ deg$^2$). The image has been smoothed with a $\sigma = 10''$ Gaussian to enhance the appearance of the diffuse emission. The subclusters, CygA and CygNW, have been modeled as spherically symmmetric.}
    \label{fig:simulation_image}
\end{figure}

\subsection{Instrument simulator}

We use the instrument simulator \texttt{MARX} \citep{dav12} to simulate the \textit{Chandra} observations. We simulate each ObsID with matching aspect solution, exposure, detector type and detector offsets. The start time of the simulated ObsIDs is matched with that of the actual ObsIDs to correctly set the effective area at the time of observation. The \texttt{MARX} tool \texttt{marx2fits} is then used to convert simulation files to a FITS event file. In order to reproject all the event files to a common tangent point, we use the CIAO tool \texttt{reproject\_obs} that takes care of the offsets between various ObsIDs and modifies the pixel coordinates accordingly. Finally, we produce counts images and exposure maps of each reprojected ObsID, and add them up using the tool \texttt{flux\_obs}. We set the \texttt{detsubsys} parameter of \texttt{mkinstmap} to \texttt{UNIFORM} in order to model the lack of spatial variation of quantum efficiency in \texttt{MARX} simulations. These datasets were then used for all subsequent analysis.

\begin{figure*}
    \centering
    \includegraphics[width=7in,height=7in, keepaspectratio]{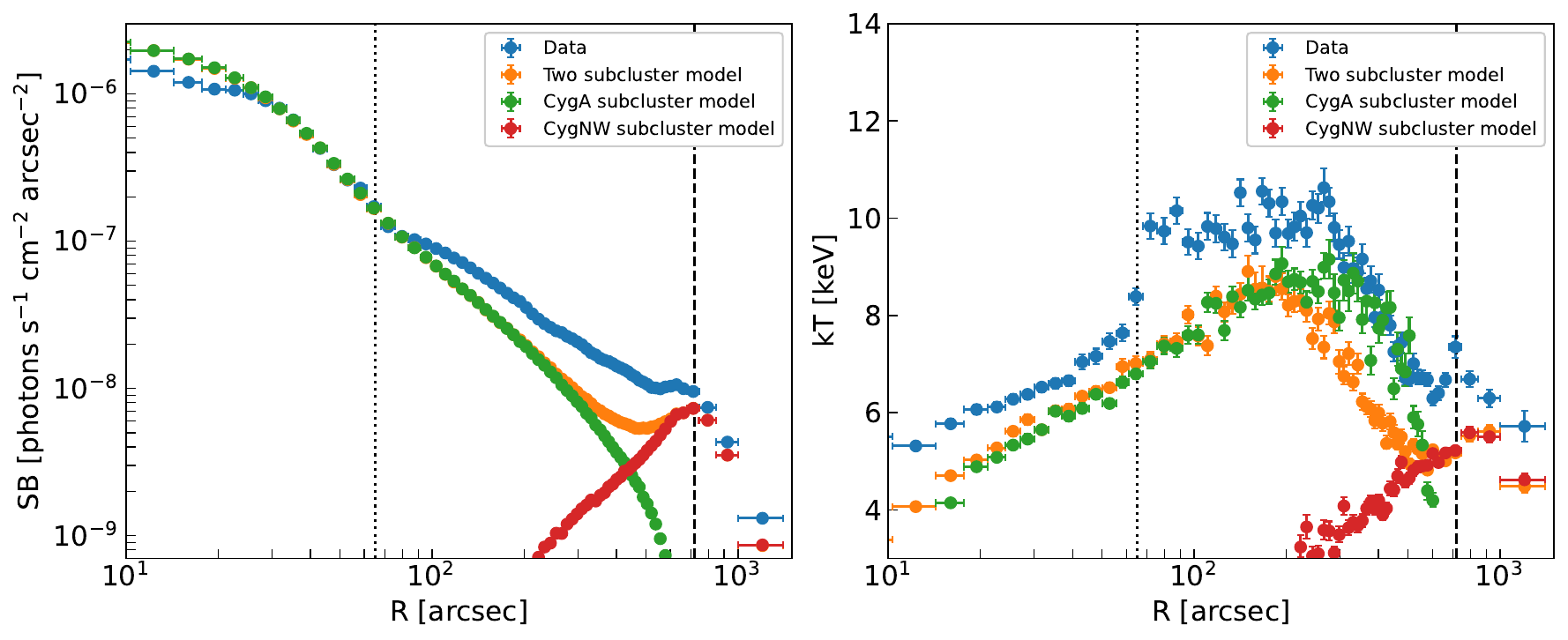}
    \caption{\small \emph{Left}: Surface brightness along the merger region for our two subcluster model, CygA only model and CygNW only model. \emph{Right}: Temperature profile along the merger region for our two subcluster model, CygA only model and CygNW only model. The presence of two clusters of different temperature alters the net temperature in our two subcluster model. The \textit{Chandra} SB and temperature profiles along the merger direction have been shown for comparison.}
    \label{fig:prop_compare}
\end{figure*}

\subsection{Input thermodynamic profiles} \label{thermodynamic_profiles}

To define the input properties of the models, we fit Equations  \ref{equation_nenp}, \ref{equation_kT} and \ref{equation_z} to the thermodynamic properties in the non-merger regions of both the CygA and CygNW subclusters. However, these input profiles are in 3-dimensional space and hence can not be directly fitted to the projected profiles shown in section \ref{profiles}. As such, we first convert the \texttt{APEC} normalization $n_en_pdV$ from spectral fits to $n_en_pdl$ by dividing out the extraction area. Equation \ref{equation_nenp} can then be integrated along the line of sight and fitted to this new profile. Similarly, the 3D temperature and metallicity profiles in equations \ref{equation_kT} and \ref{equation_z} can be converted to projected 2D profiles by:

\begin{equation}
    Y_{proj}(r) = \frac{\int \omega Y(r)dV}{\int \omega dV}
\end{equation}
where $Y(r)$ can be a 3D profile like temperature or metallicity, $Y_{proj}(r)$ their 2D analog and dV the volume along line of sight. The weighting function $\omega$ is assumed to be proportional to emissivity of each gas element along the line of sight $\omega \propto n_en_p\sqrt{T}$ \citep{sar86}. These profiles can now be easily fitted to the projected temperature and metallicity profiles along the non-merger direction of CygA and CygNW. These fits are shown in Figure \ref{fig:marx_fit}. As there is evidence of contamination of CygNW in the density profile beyond 600 arcsec (see \S\ref{density}), we limit the CygA fits below this distance. The CygNW profiles are calculated up to the edge of field of view. The fitted parameter values obtained from such fits are reported in \S\ref{best_fit}. These parameter values completely define the input profiles and can be used to start the simulation. 

\subsection{Simulation results} \label{analysis}


The exposure-corrected image from the simulation is shown in Figure \ref{fig:simulation_image}. The two subclusters overlap in the region in between which suggests that the surface brightness and temperature profile in this region should be different from individual subcluster profiles. This effect is shown in Figure \ref{fig:prop_compare} where a comparison of SB and temperature profiles is shown. These profiles have been calculated by extracting simulated counts along the overlapping region using the same annular bins discussed in section \ref{regions}. The temperature profiles are calculated by fitting a single temperature absorbed \texttt{APEC} model. We use the \texttt{wabs} model for absorption and keep the column density fixed at $n_H = 4.14 \times 10^{21}$ atoms cm$^{-2}$. We obtain fits with \texttt{cstat/dof = 38543/30167} in the outermost bin.

We note that the surface brightness in the `Two subcluster model' is simply a sum of the surface brightness of two individual subclusters. The surface brightness from the `Two subcluster model' is lower than that derived from the \textit{Chandra} data by a factor of a few (See Figure \ref{fig:prop_compare}). This may mean that there is additional gas in the merger region than predicted by the input thermodynamic profiles that were obtained by fitting density profiles on the non-merger side of the two subclusters (Figure \ref{fig:marx_fit}). Furthermore, there maybe additional heating due to the merger. These two effects explain why the model surface brightness does not match the surface brightness from the data in the merger region.

On the other hand, the temperature profile for the two subcluster scenario is not the average of the temperatures from two clusters as one may trivially expect. This is because the net temperature profile for the two subcluster model is determined by the underlying 3D temperature and density profile. We find that the two subcluster temperature profile agrees with the single subcluster profiles near the core of the subclusters. However, it is lower in the overlapping region than what one may expect when only the CygA subcluster is present. This is a non-trivial effect due to the presence of two non-interacting gaseous spheres. 

\begin{figure*}
    \centering
    \includegraphics[width=7in,height=7in, keepaspectratio]{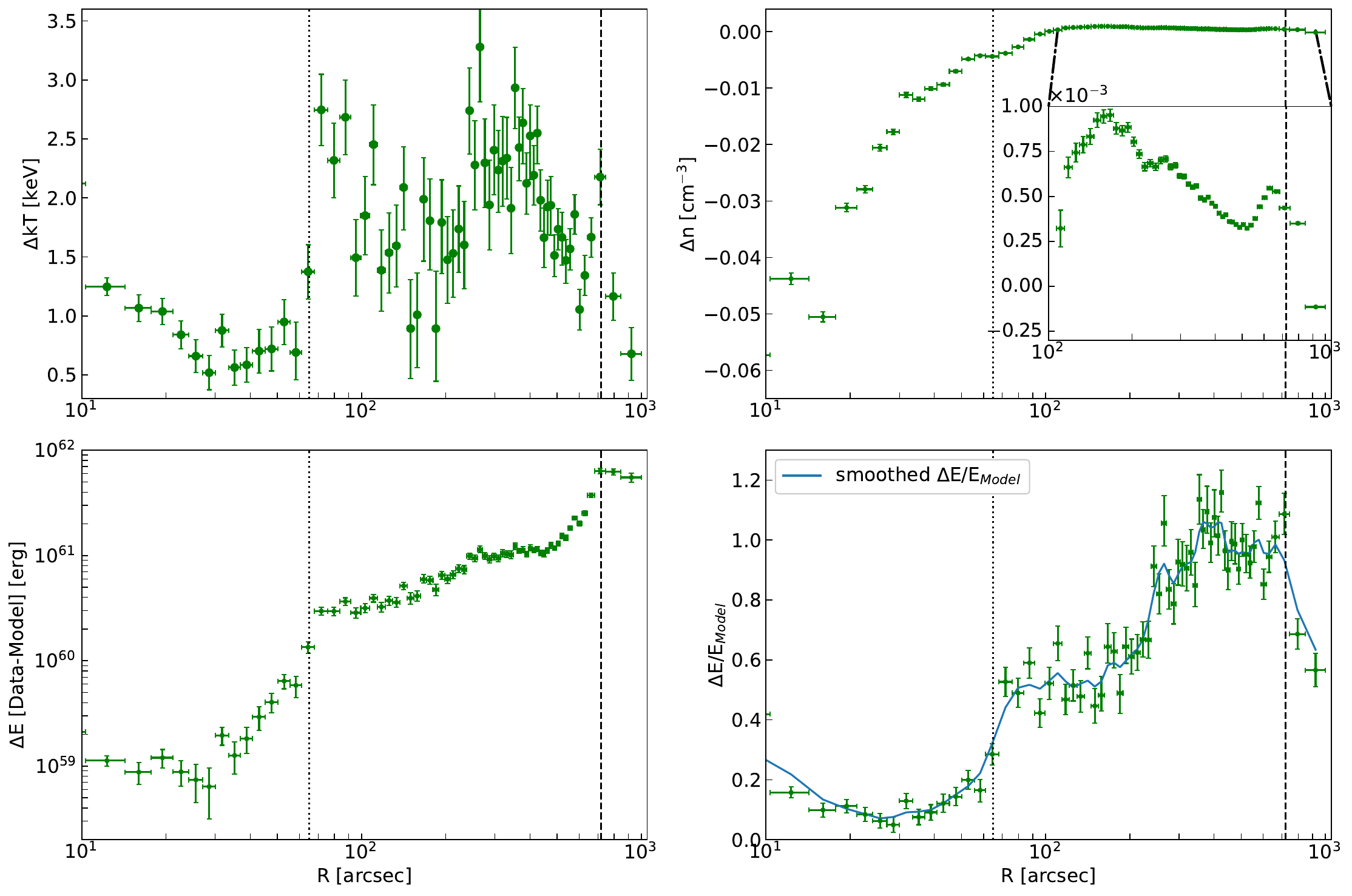}
    \caption{\small \emph{Upper Left:} Excess temperature along the merger direction. \emph{Upper Right:} Excess total density along the merger region. \emph{Lower Left:} Excess energy along the merger direction. \emph{Lower Right:} Excess energy along the merger direction divided by the energy predicted by the two subcluster model. The blue line shows the gaussian smoothed excess energy. The dotted line shows the location of the cocoon shock while the dashed line shows the location of the CygNW subcluster.}
    \label{fig:excess_energy}
\end{figure*}

\section{Energy excess in the merger region} \label{excess_energy}

From \S\ref{analysis}, it is clear that the interaction in the merger region creates a substantial change in the properties of the gas. We investigate this interaction by subtracting the two subcluster model temperature profile from the \textit{Chandra} temperature profile along the merger direction. This excess temperature is shown in Figure \ref{fig:excess_energy}. We further show the excess density in the merger region by subtracting the density in the two subcluster model from the \textit{Chandra} density profile. With both excess temperature and excess density, we can derive the excess energy in the merger region using the following equation:

\begin{equation}
    \Delta E = \frac{5V}{2} \left\{(nkT)_{\textrm{Data}} - (nkT)_{\textrm{Model}}\right\}   \label{energy}
\end{equation}
Here $(nk\Delta T)_{\textrm{Data}}$ is obtained by multiplying density and temperature from the \textit{Chandra} data. $(nk\Delta T)_{\textrm{Model}}$ is similarly obtained by multiplying density and temperature from the model. $V$ corresponds to effective volume along the 
line of sight (See Eq. \ref{volume}). We note that assuming a spherical geometry is a major simplification and the true geometry may differ significantly. Such information about the geometry can only be obtained using numerical simulations assuming different geometries and comparing to observation. This is beyond the scope of this paper. Assuming such a simplification however, the excess energy from the merger is shown in Figure \ref{fig:excess_energy}. It is hard to visualise any feature from this figure however. Hence, we divide this excess energy by the thermal energy predicted by our two subcluster model and show it in the same figure. We also show a gaussian smoothed $\Delta E/E_{\textrm{Model}}$ curve for better visualisation of the fluctuations.

In Figure \ref{fig:excess_energy}, we see several interesting features. Firstly, there is an overall temperature, density and energy excess over the entire merger region. Secondly, there are multiple fluctuations in both temperature and in the $\Delta E/E_{\textrm{Model}}$ plot. These fluctuations are associated with the prominent filamentary structures seen in the \textit{Chandra} image. The positive excess everywhere is expected from detailed numerical simulation of the CygA system \citep{hal19}, but smaller scale temperature fluctuations observed in Figure \ref{fig:excess_energy} are not easily explained by the merger. We provide a hypothesis for the cause of these fluctuations and model the $\Delta E/E_{\textrm{Model}}$ curve in the next section.  

\subsection{Merger and AGN contribution to excess energy} \label{relative_contribution}

The excess energy can be due to two different possible sources. A numerical simulation of the Cygnus A merging system \citep{hal19} shows that a smooth increase of temperature in the merger region is expected. This implies that the merger can partly explain the energy excess we see in the merger region. We model such a smooth increase with a simple 2nd-order polynomial expression. We assume that the effect of the merger is negligible at the cocoon shock as any energy increase in this region is likely to be dominated by the current epoch of AGN activity.  The relative energy increase due to the merger can thus be modeled as: 

\begin{equation}
\frac{\Delta E_{\textrm{Merger}}}{E_{\textrm{Model}}} (r) = a (r - r_{\textrm{cocoon}})^2 + b(r - r_{\textrm{cocoon}})
\label{equation_delkT_merger}
\end{equation}

The observed energy fluctuations however are not easily explained by the merger. We hypothesize that these fluctuations are caused by previous outbursts from the central AGN. There is evidence that the power injected by the central AGN is significantly higher than the core cooling luminosity (See Fig. 8 of \citealt{Uber23}). Excess power from outbursts may thus leave their imprint in the ICM in the form of energy fluctuations. We model these outbursts with a set of gaussians on top of the smooth temperature increase due to the merger. 
\begin{table*}
\caption{Properties of merger and outburst signatures. $\Delta E/E_{\textrm{Model}}$ is the amplitude of the outburst. Radius is distance of the peak of the outburst from the center of CygA subcluster. FWHM is the full-width at half maximum of the outburst. $t_{age}$ is the age of the outburst. $\Delta t$ is the outburst timescale. $E_{out}$ is energy output from the outburst. $P_{out}$ is the power generated by the outburst. The merger excess energy is modeled by a polynomial and we report the output energy, $E_{out}$. The 1st gaussian peak is residual from CygNW (see \S\ref{excess_energy}).}
\centering
\begin{tabular}{c c c c c c c c c} 
\hline
\hline
Model component & $\Delta E/E_{\textrm{Model}}$ & Radius & FWHM & $t_{age}$       & $\Delta t$            & $E_{out}$               & $P_{out}$  & Remarks \\
                                  &                         & (kpc)    & (kpc)     &  (Myr)   &  (Myr)      & ($10^{61}$ ergs)  & ($10^{46}$ erg s ${^{-1}}$)    &  \\
\hline
1st Gaussian & $0.40 \pm 0.02$ & $90 \pm 2$ & $41 \pm 4$ & $61.8 \pm 0.4$ & $26 \pm 4$ & $1.2 \pm 0.1$ & $1.5 \pm 0.3$ & 1st outburst \\
2nd Gaussian & $0.14 \pm 0.03$ & $133 \pm 4$ & $36 \pm 10$ & $88 \pm 3$ & $22 \pm 7$ & $0.4 \pm 0.2$ & $0.6 \pm 0.3$ & Possible 2nd outburst \\
3rd Gaussian & $0.17 \pm 0.04$ & $436 \pm 13$ & $103 \pm 36$ & $278 \pm 9$ & $69 \pm 24$ & $1.5 \pm 0.7$ & $0.7 \pm 0.4$ & Possible 3rd outburst \\
\hdashline
4th Gaussian& $1.07 \pm 0.03$ & $746 \pm 8$ & $308 \pm 33$ & --- & --- & --- & --- & CygNW residuals \\ 
Polynomial &  ---   &  ---   &  ---   &  ---   &  ---   &   $32 \pm 3$  & --- & Merger\\
\hline
\end{tabular}
\label{delkT_fit_properties}
\end{table*}

In Figure \ref{fig:chandra_delkT_fits}, we show such toy models with different numbers of gaussian-shaped outbursts. Multiple outburst components are necessary to reduce the fit residuals, as shown in the figure. However, we take caution to not overfit the data with more components than what is necessary. To do this, we use a statistical technique called Bayesian Information Criterion (BIC). The Bayesian information criterion can be written as:  

\begin{equation}
BIC = \chi^2 + k \log(N)
\label{equation_bic}
\end{equation}

Here $k$ is number of parameters, $N$ is number of data points and $\chi^2 = \sum_{i}^{} (n_i-s_i)^2/\sigma_{i}^{2}$, where $n_i$ is the $i$th data point, $s_i$ is the $i$th model value and $\sigma_{i}$ is the error on $i$th data point. The BIC penalises addition of parameters with the $k\log(N)$ term and optimizes the number of parameters necessary to reduce the residuals. The number of components for which the BIC is minimum is considered the best model. We note the BIC values of our models in Table \ref{delkT_fit_table}. The mean values and their errorbars are calculated from BIC values of 1000 realisations that assumes the $\Delta E/E$ values are gaussian distributed around their mean. It is clear that four gaussians along with the merger component leads to the minimum BIC value. But the BIC values for `Polynomial + 3 gaussians' and `Polynomial + 4 gaussians' models are within 1$\sigma$ error of the BIC value for `Polynomial + 2 gaussians' model. Thus, we note these two components as possible outbursts in Table \ref{delkT_fit_properties}. We also derive properties of all outburst components for completeness sake. Including more outburst components leads to an increase in the BIC value and thus they are not favored. 

We note that the excess energy bump at the CygNW position is likely due to residual emission from CygNW rather than any effect due to the AGN. CygNW is an extremely diffuse cluster undergoing tidal disruption and hence the true geometry of CygNW is probably different from the spherical geometry assumed in the simulation. 

We multiply the relative excess energy with the energy predicted by our two subcluster model to get the absolute excess energy due to the merger and outbursts. We calculate total excess energy of each component by adding the excess energy value at every radial bin. We find that the merger is adding energy up to $(3.2 \pm 0.3) \times 10^{62}$ erg into the ICM. The total energy injected by all the outbursts is $(3.1 \pm 0.7) \times 10^{61}$ erg, which is $\sim$10\% of the merger energy. Furthermore, we calculate the full-width at half maximum (FWHM) for each gaussian component and derive outburst timescale as $\Delta t = 2 \times \textrm{FWHM}/c_s$, where $c_s$ is local sound speed. The age of each outburst is calculated by noting the distance of the fitted gaussian peak from the central AGN and integrating the time taken by a sound wave to reach that distance assuming that the density and temperature varies according to the profiles discussed in \S\ref{profiles}. Finally, we calculate the power of each outburst by dividing its energy output by the corresponding outburst timescale, $\Delta t$. All the derived properties for each outburst are reported in Table \ref{delkT_fit_properties}. The mean values and their errorbars are calculated by assuming that fit parameter values are gaussian distributed around their best-fit values and calculating 1000 realisations of the fit parameters. 

The results in Table \ref{delkT_fit_properties} have interesting implications regarding the history of Cygnus A. The derived ages of the outbursts imply that Cygnus A has been active for the past $\sim 300$ Myr with duty cycle of $\sim 30-200$ Myr. These values of duty cycles are similar to the values determined for AGN-induced cavity systems in other cluster cores (for reviews, see \citealt{mcn07,mcn12}). The measured ages in Table \ref{delkT_fit_properties} suggest that the duty cycle between outbursts could have reduced over time.  We do not see any temporal evolution of outburst timescale $-$ older and younger events have similar outburst periods within $2\sigma$ errors. It does seem though that the power of the outbursts may have increased with time. Such a scenario is consistent with the result found by \cite{cho12}. However, we have only 3 possible outburst candidates so any evidence is marginal at best. Also, if the 2nd and 3rd Gaussian in Table \ref{delkT_fit_properties} are real, then we can say that CygA has been consistently active for the past $\sim 300$ Myr.

\begin{table}
\caption{Bayesian Information Criterion (BIC) values for different models.}
\centering
\begin{tabular}{c c} 
\hline
\hline
Model & BIC \\
\hline
Polynomial only & 670 $\pm$ 50\\ 
Polynomial + 1 gaussian & 500 $\pm$ 40\\
Polynomial + 2 gaussians & 240 $\pm$ 30\\
Polynomial + 3 gaussians & 220 $\pm$ 30\\
Polynomial + 4 gaussians & 210 $\pm$ 30\\
\hline
\end{tabular}
\label{delkT_fit_table}
\end{table}

\begin{figure*}
    \centering
    \includegraphics[width=8.5in,height=8.5in, keepaspectratio]{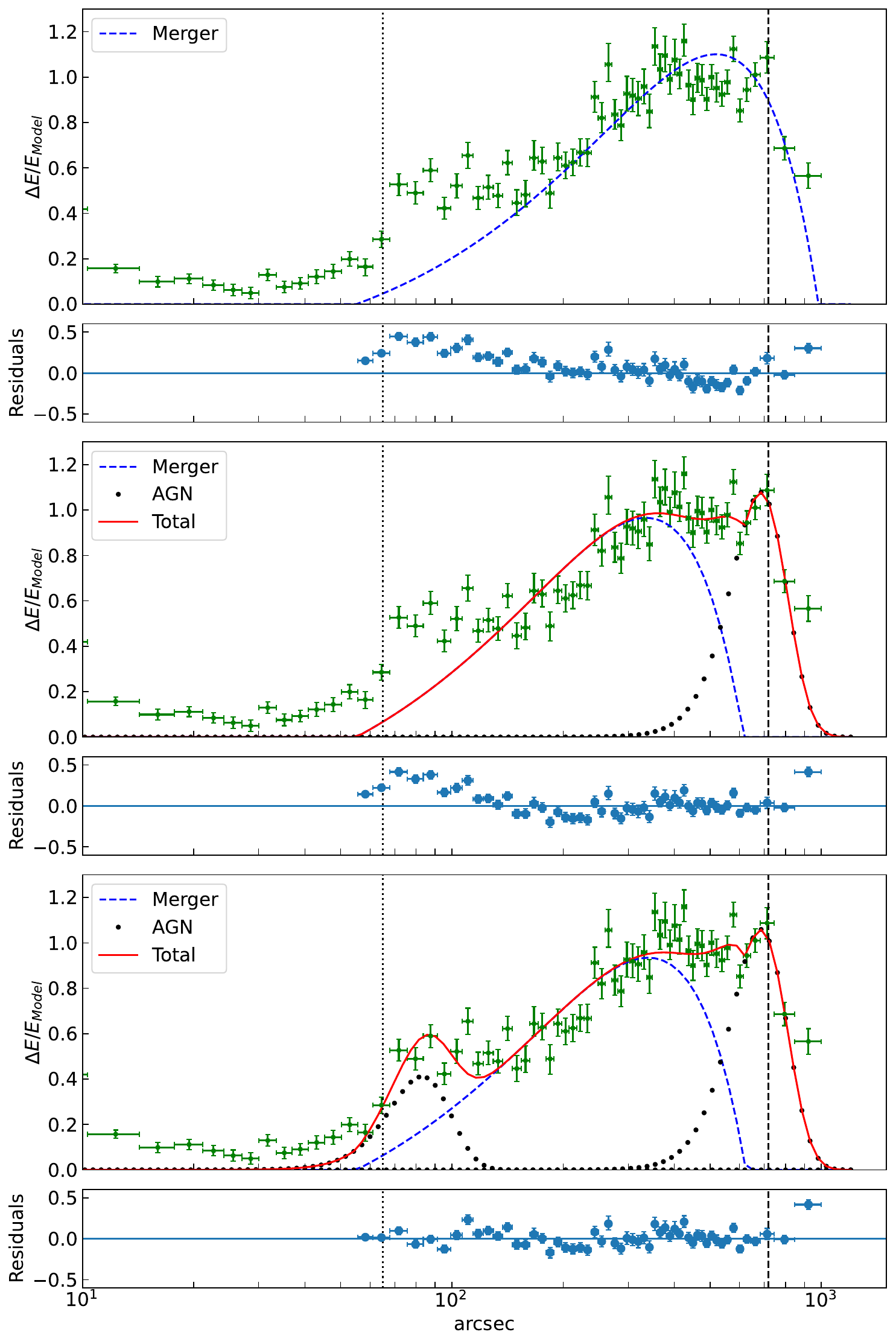}
    \caption{\small \emph{Top:} Residual temperature fitted with only a 2nd order polynomial. This represents the merger heating. \emph{Middle:} Fit with polynomial + 1 gaussian. AGN outbursts are approximated as gaussians. \emph{Bottom:} Fit with polynomial + 2 gaussians. The dotted and dashed lines show the location of the cocoon shock and CygNW cluster.}
    \label{fig:chandra_delkT_fits}
\end{figure*}

\begin{figure*}
    \ContinuedFloat
    \captionsetup{list=off}
    \centering
    \includegraphics[width=8.5in,height=8.5in, keepaspectratio]{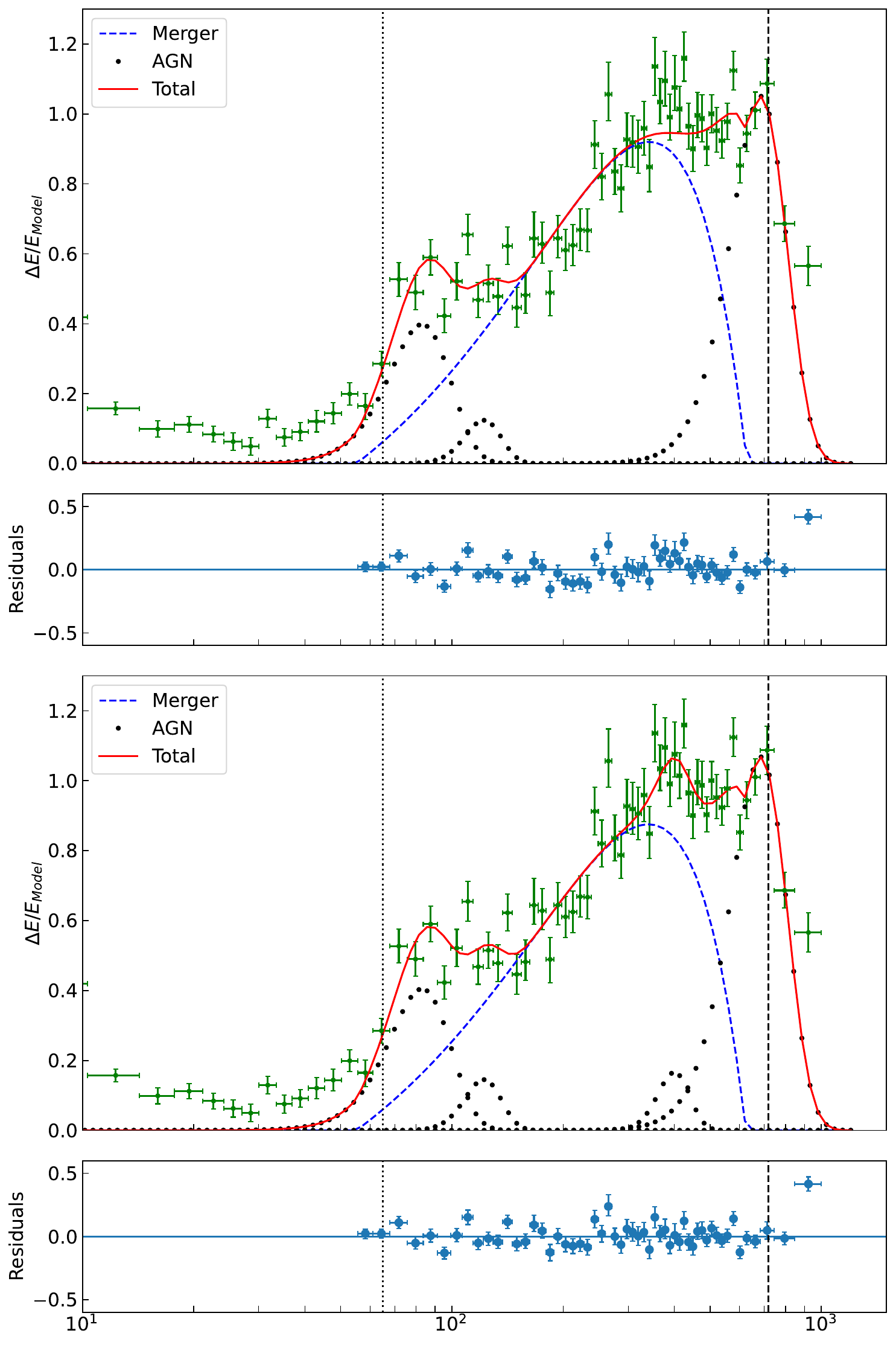}
    \caption{\small (contd.) \emph{Top:} Fit with polynomial + 3 gaussians. \emph{Bottom:} Fit with polynomial + 4 gaussians.}
    \label{fig:chandra_delkT_fits_contd}
\end{figure*}

With the exception of the most recent episode associated with the cocoon shock, the power for other two outbursts does not show any significant evolution over the past $\sim 300$ Myr. The powers quoted in Table 3 are lower limits on the outburst energies since they assume that the underlying large-scale temperature increase is entirely due to the merger. Although well motivated by numerical simulations, the shape and amplitude of the merger parametric model as a function of radius is somewhat arbitrary. For instance, if we assume that there is little merger contribution beyond a few hundred arcsec, the derived outburst energy values will be higher. Numerical simulations will be needed to explore such possibilities and are beyond the scope of this paper. 

In this study, we used the thermodynamic profiles from the non-merger side of CygA and CygNW to create our simulated subclusters. In doing so, we assumed the non-merger side of CygA does not contain AGN heating. We make this assumption as we do not see signatures of outbursts on the non-merger side with the present data quality. Nevertheless, the AGN's outbursts may also be heating the ICM on the non-merger side. In that case, the derived outburst energy in this section is a lower estimate than the actual total outburst energy.

\section{Conclusions} \label{summary}

In this work, we have used 2.2 Ms \textit{Chandra} and 40 ks \textit{XMM-Newton} data to study the Cygnus A merging system. Our primary objective was to study the hot ICM in between the CygA and CygNW subclusters and investigate the mechanisms responsible for heating the ICM. We resolved the X-ray features in the ICM from kpc scale to Mpc scale and our results can be summarised as follows:

\begin{itemize}
    \item The Cygnus A merger system shows diffuse emission extending over a Mpc. The merger is between two distinct subclusters that we call CygA and CygNW. The merger region shows enhanced surface brightness and filamentary structures. In previous low resolution studies \citep{mar99,sar13}, this region has been associated with higher temperature due to the merger.
    \item We confirm the results seen in previous analysis finding a significant temperature enhancement in the merger region relative to the surrounding cluster gas outside this region. With a much deeper exposure, however, we also notice temperature fluctuations of the order of $0.5-2.5$ keV on top of the underlying profile. These temperature fluctuations correspond spatially with the filamentary structures seen in the surface brightness maps. 
    \item In order to separate the contribution of the merger from any additional heating mechanisms, we have constructed a model of CygA and CygNW subclusters based on their locations and physical properties as determined outside the merger region. The subclusters in our model do not interact with each other, providing a baseline scenario with respect to which any heating in the merger region of actual data can be compared. The temperature profile in our two subcluster model differs substantially from the case of a single subcluster due to the presence of multi-phase gas. We also conclude that the merger region has more gas than that predicted by smooth hydrodynamic profiles.
    \item The temperature difference between the actual data and our model, $\Delta kT$, shows a positive excess at all radii from the center of CygA. The temperature excess varies between $0.5-2.5$ keV with significant temperature fluctuations. The derived energy excess also shows such fluctuations and when smoothed, the energy excess suggests the presence of ripples on top of a smooth underlying profile. We hypothesise that this excess energy can be best explained by heating due to a merger shock and a series of AGN outbursts that have propagated a distance of few hundred kpc from the center of Cygnus A.  
    \item The energy excess is best modeled by a set of two gaussians representing AGN outbursts and a smooth polynomial representing the ongoing merger. There is weaker evidence that there may have been two more outbursts in the history of Cygnus A. If these additional outbursts are indeed present, Cygnus A has been active over the past 300 Myr with a duty cycle of $30-200$ Myr. The smooth merger component, on the other hand, injects up to $\sim$$10^{62}$ erg into the ICM. Furthermore, the total outburst energy can be $\sim10\%$ of the merger energy or higher. Assuming our model of merger heating + AGN outburst is correct, this suggests that AGN outbursts can be a significant driver of ICM heating over the lifetime of a cluster.
    
\end{itemize}

\section*{Acknowledgements}

 AM thanks Dr. John Zuhone (Harvard-Smithsonian CfA) for helping to set up the simulation in this paper. The authors thank Dr. Daniela Huppenkothen (SRON) for valuable advice on statistical analysis. The authors also thank the anonymous referee for their comments during the review process. Plots were generated using \texttt{MATPLOTLIB} \citep{hun07} and \texttt{APLPY}\footnote{\href{http://aplpy.github.com}{http://aplpy.github.com}}. This research also made use of \texttt{NUMPY} \citep{vdw11}, \texttt{SCIPY} \citep{jon01} and \texttt{ASTROPY} \citep{ast13}.  

\section*{Data Availability}

The \textit{Chandra} and \textit{XMM-Newton} data used in this article are publicly available. All the codes, simulation data and other materials used to generate the results of this paper are publicly available in Zenodo \citep{maj24_1,maj24_2}.

\input{draft.bbl}


\appendix
\section{Best-fit parameters} \label{best_fit}

The best-fit parameters for density, temperature and metallicity profiles, as discussed in Eqs. \ref{equation_nenp}, \ref{equation_kT} and \ref{equation_z} are reported in Tables \ref{density_param}, \ref{temperature_param} and \ref{metallicity_param}. The fits are insensitive to some parameter values, which were therefore kept fixed while fitting. These values have been used to create the simulation results in \S\ref{Simulation}. We caution the readers that some parameters are degenerate and it is possible to obtain slightly different parameter values based on the initial guess. 

\begin{table*}
\caption{Best-fit parameters for density profile. Parameters that were held fixed while fitting are quoted without errors.}
\begin{tabular}{c c c c c c c c c c c} 
\hline
\hline
Subcluster & $n_0$                  & $r_c$         & $r_s$         & $\alpha$       & $\beta$                        & $\epsilon$                &    $\gamma$     &$n_{02}$               & $r_c$    &   $\beta_2$\\
           & 10$^{-2}$ cm$^{-3}$    & (kpc)         & (kpc) &               &         &     &       &   10$^{-2}$ cm$^{-3}$      & (kpc) &     \\
\hline

Cygnus A   & $2.7 \pm 0.2$          &  $31 \pm 2$   &  250  &  10$^{-38}$   &   $0.419 \pm 0.004$  &   0.7  &   3   & $0.44 \pm 0.08$       &  70      &  2      \\
Cygnus NW  & $2.29 \pm 0.07$         &  $116 \pm 16$ & 10000        &   10$^{-8}$    &   $0.38 \pm 0.03$              & 30000  &   3     &  0.55                    &  15       &    10 \\

\hline
\end{tabular}
\label{density_param}
\end{table*}

\begin{table*}
\caption{Best-fit parameters for temperature profile.}
\begin{tabular}{c c c c c c c c c } 
\hline
\hline
Subcluster & $T_0$                  & $r_t$ & $a$ & $b$ & $c$ & $T_{min}/T_0$                  & $r_{cool}$             & $a_{cool}$    \\
           & (keV)                  & (kpc) &     &     &     &                     & (kpc)                  & \\
\hline

Cygnus A   &  $9.96 \pm 0.07$   & 390   &   $-0.264 \pm 0.004$   &  $9 \pm 1$   &   $1.37 \pm 0.03$  &  5       &       18      &       5       \\
Cygnus NW   &  $3.5 \pm 0.7$   & 310   &   $0.19 \pm 0.03$   &  900   &   $0.59 \pm 0.08$  &  6       &       670      &       190       \\
\hline
\end{tabular}
\label{temperature_param}
\end{table*}

\begin{table*}
\caption{Best-fit parameters for metallicity profile.}
\begin{tabular}{c c c c} 
\hline
\hline
Subcluster & A                  & B                     &           C \\
\hline
Cygnus A   & $0.187 \pm 0.006$  & $0.0065 \pm 0.0001$   &          $-0.48 \pm 0.01$ \\
Cygnus NW  & $0.09 \pm 0.01$    &   -0.16               &       -1    \\
\hline
\end{tabular}
\label{metallicity_param}
\end{table*}


\bsp	
\label{lastpage}
\end{document}